\documentstyle[11pt]{article}
\begin{document}


\title{Hsiao-Code Check Matrices and Recursively Balanced Matrices}

\author{  Li Chen \\
Department of Computer Science and Information Technology\\
University of the District of Columbia\\
Washington, DC 20008 \\
lchen@udc.edu
}

\maketitle

\begin{abstract}

The key step of generating the well-known Hsiao code is 
to construct a $\{0,1\}$-check-matrix 
in which each column contains the same odd-number of 1's and each row contains
the same  number of 1's or differs at most by one for the number of 1's. 
We also require that no two columns are identical in the matrix.   
The author solved this problem in 1986 by introducing a type of 
recursively balanced matrices. However, since the paper was published in Chinese, 
the solution for such an important problem was not known by international
 researchers in coding theory.
In this note, we focus on how to practically generate
the check matrix of Hsiao codes. We have modified the original algorithm
to be more efficient and effective. We have also corrected an error in
algorithm analysis presented in the earlier paper. The result shows
that the algorithm attained optimum in average cases if a
divide-and-conquer technique must be involved in the algorithm.

\end{abstract}


\section {Introduction}

Error-detections and
corrections are required for computer main memory and secondary storages.
In recent years, trusted computing and computing
reliability have become more and more important in theory and practice. Data recovery is at
the center of concerns. Error-corrections codes, especially Hamming codes and
Hsiao codes are still essential to this type of technology \cite{Sta}\cite{Lal} .

The well-known Hsiao code is the most energy saving Hamming Codes \cite{Hsi}\cite{Lal}\cite{Gan}\cite{GBT04}\cite{GBT05}. 
It has been widely used
in memory fault tolerance for more than thirty decades. As a type of SEC-DED codes, 
i.e. single error
correction and double-error detection codes,
Hsiao codes attained the optimal in minimum odd-weight of columns. Even though there was 
an observation indicating that cellular automata-based codes might be better than
Hsiao codes in terms of check bits for cellular automata --a future computing device
\cite{Gan}\cite{Cho} . 
However, Hsiao code is still the most efficient code used 
in industry \cite{Lal}\cite{GBT05}.

After a code is designed, the most important task is to find its check matrix. 
The author introduced an algorithm that recursively constructed the check-matrix to Hsiao
codes   in 1986 \cite{Che86}. Unfortunately, 
the published paper was written in Chinese and the title of the paper did not mention 
the SEC-DED code.  This algorithm was not known by international researchers in coding theory.

In this paper, we first explain the algorithm described in \cite{Che86} where it was
addressed in an abstract and in a brief manner.  Then, we modified some steps
to show a more
detailed algorithm. Finally, we corrected an error in the algorithm analysis
of \cite{Che86} and provided a more precise time analysis for general cases.
Theoretically, both original algorithm and the modified algorithm
are very efficient and fast in terms of complexity theory. 
We have proved that the modified algorithm is optimal in average cases.
However, in practice, we may find an even more efficient algorithm.

\section {Major Steps of Generating Check Matrix ${\bf H}$ for Hsiao Code }

The definition of Hsiao code is a type of SEC-DED codes whose
check matrix $\bf H$ defined on $GF(2)$ satisfies:
\begin{itemize}
\item [(1)] Every column contains an odd number of 1's.
\item [(2)] The total number of 1's reaches the minimum.
\item [(3)] The difference of the number of 1's in any two rows is not greater than 1.
\item [(4)] No two columns are the same.
\end{itemize} 
\noindent Therefore, Hsiao called this code an optimal minimum odd-weight-column SEC-DED code.

Now, we will discuss how to generate $\bf H$.
Assume that we want to generate the SEC-DED Code with $k$
information (data) bits, first according to the general   requirement of Hsiao codes \cite{Hsi},

\begin{equation}
    R \ge 1 + \log_{2}{( k + R )}
\end{equation}

\noindent From this, we can determine $R$. After $R$ is determined, 
we define $\Delta(R, J, m)$ a $\{0,1\}$-type
$R\times m$ matrix with column weight $J$, $0\le J\le R$. No two columns are 
the same in $\Delta(R, J, m)$.

\begin{equation}
      \Delta(R, 2\cdot i +1, C_{R}^{2\cdot i +1}),  i = 0,1,..., I-1.
\end{equation}

\noindent Where,  $C_{R}^{2\cdot i +1}$  =
$\left ( \begin{array}{c}
          R\\
          2\cdot i +1
          \end{array}
          \right)$ is the combinatorial number. Assume,

\begin{equation}
  \Sigma_{i=0}^{I-1} C_{R}^{2\cdot i +1} \le (k+R) \le \Sigma_{i=0}^{I}  C_{R}^{2\cdot i+1}. 
\end{equation}

Let $m=(k+R)- \Sigma_{i=0}^{I-1} C_{R}^{2\cdot i +1}$. 
We can see that the most important task is to generate

\[   \Delta(R, 2\cdot I +1, m) \]

\noindent because the other cases are the special cases of it.

\noindent Thus,
\begin{equation}
  H = [\oplus_{i=0}^{I-1} \Delta(R, 2\cdot i +1, C_{R}^{2\cdot i +1})] 
  \oplus  \Delta(R, 2\cdot I +1, m), 
\end{equation}
\noindent where $\oplus$ is to union these arrays together, horizontally.

\section {The Algorithm of Generating  $\Delta(R, J, m)$   }

According to Section 2, it is not difficult to see that obtaining the
matrix $\Delta(R, J, m)$ is the key
to the check matrix ${\bf H}$, especially when $m\ne C_{R}^{2\cdot i + 1}$.

Let's first define some notations:
\begin{itemize}
\item [(a)] {\bf $\Delta(R, J, m)$:} a $\{0,1\}$-type 
                 $R\times m$ matrix with column weight $J$, $0\le J\le R$.
\item [(b)] {\bf $<t>(m)$:} a $\{0,1\}$-type $1\times m$ matrix in which 
               every component is $t$. For example, $<1>(2) = (1,1)$.

\item [(c)] {\bf $\circ$:} the matrix up-down union operator, i.e. $N\circ M =
             \left [ \begin {array}{c} a\\b
                   \end{array} \right ] $

\item [(d)] {\bf  $\oplus$:} the matrix left-right union operator, i.e. $N\oplus M = [N M]$

\item [(e)] {\bf  $\overline{\oplus}$:} $N\overline{\oplus}M$ is to place the matrix $M$ 
             upside down, do ``$\oplus$,'' and then move the rows that contain more 1's to the top of the matrix.

\item [(f)] {\bf  $L-condition$:} $\Delta(R, J, m)$ is said to satisfy the condition if $0 \le J \le R$ $\&$ $0 \le m \le C_{R}^{J}$.
 
\item [(g)] {\bf Matrices with equal-weight-columns/rows:} Matrices have equal-weight columns
             and quasi-equal-weight rows. In other words, in these matrices, each
             column has the same number of 1's and
             each row has almost the same number of 1's.
             (The difference of the numbers of 1's is less than 1.) 
             We will also call such a matrix a
             (weighted) Balanced matrix.
\end{itemize}

We also define the following special $\Delta(R, J, m)$ that will keep the rows with more 1's at the
top of the array:
\begin{itemize}
 
\item [(1)] If $m=0$, $\Delta(R, J, m)= \emptyset $.

\item [(2)] If $J=0$, $\Delta(R, J, m)= (0,...,0)^{T}$.

\item [(3)] If $J=R$, $\Delta(R, J, m)= (1,...,1)^{T}$.

\item [(4)] If $m=1$, $\Delta(R, J, m)= (1,...,1,0,...,0)^{T}$, where the number of 1's is $J$.

\item [(5)] If $J=1$, $\Delta(R, J, m)= \left ( \begin {array}{cccc} 1 & 0 & ...&0 \\
                                                             ...& ...&  &...\\
                                                             0  & ...& 0 & 1 \\
                                                             0   & ...&  & 0 \\
                                                             ...& ...&  &...\\
                                                              0   & ...&  & 0
                   \end{array} \right ) $, there are $m$ 1's in the matrix.

\item [(6)] If $J=R-1$, $\Delta(R, J, m)= \left ( \begin {array}{cccc} 1 & 1 & ...&1 \\
                                                             ...& ...&  &...\\
                                                             1  & ...& 1 & 1\\
                                                             0   & 1 &... & 1\\
                                                             1   & 0 &... & 1 \\
                                                             ...& ...&  &... \\
                                                              1   &1 & ... & 0
                   \end{array} \right ) $, there are $m$ 1's in the matrix.
 
\end{itemize}

\noindent The above matrices are called the ending-states. 
So we can assume that $2\le J\le R-2$ for later discussion.

Let $\Delta(R, J, m)$ satisfy the $L$-condition. We can represent $\Delta(R, J, m)$
in the following form.

\begin{equation}
   \Delta(R, J, m) = \left [ \begin {array} {cccccccc}
                              1 & ...............& 1 & 0 &...............& 0 \\

                               &\Delta_{1}(R-1, J-1, m_{1})& & &\Delta_{2}(R-1, J, m-m_{1})&

                             \end{array} \right ]  
\end{equation}

{\bf Theorem 1 }\cite{Che86}  Assume $\Delta(R, J, m)$ satisfy the $L$-condition. Let $m_{1}=
[\frac{m\cdot J}{R} + \frac{R-1}{R}]$, then
 $\Delta_{1}(R-1, J-1, m_{1})$ and $\Delta_{2}(R-1, J, m-m_{1})$ also satisfy
$L$-condition.

{\bf Theorem 2 }\cite{Che86} Suppose that $\Delta_{1}(R-1, J-1, m_{1})$ and
$\Delta_{2}(R-1, J, m-m_{1})$ are Matrices with equal-weight-columns/rows, where $m_{1}=
[\frac{m\cdot J}{R} + \frac{R-1}{R}]$.  Then,
\begin{equation}
 \Delta'(R, J, m) = [ <1>(m_{1}) \oplus <0> (m-m_{1})]
                      \circ
          [\Delta_{1}(R-1, J-1, m_{1}) \overline{\oplus} \Delta_{2}(R-1, J, m-m_{1})]  
\end{equation}
\noindent is a matrix with equal-weight-columns/rows.

The above two theorems indicate a recursive process
of generating a matrix with equal-weight columns
and quasi-equal-weight rows:
The Theorem 2 provides the union process that guarantees that each row
after the union has about the
same weight. $\overline{\oplus}$, to place the matrix on the right side of the operator
 upside down,
 is to avoid one row
containing more than one 1's
after the merge. It is the simplest mathematical way of solving the problem.
The recursive matrice in (5) is also called a recursively balanced matrix.

However, by ``placing a matrix upside down'' and then moving rows
with more 1's to the top of the matrix after merging takes more
computational time. In \cite{Che86}, the author made a mistake by simply stating that
this process is optimum $O(R \cdot m)$. This is because some of the
parts in the array may need to be
``placed upside down'' many times (say $\min\{R,m\}$). So , in the worst case,
the complexity of the procedure
given in Theorem 2 is $O(R \cdot m \cdot \min\{R,m\})$. This is still a
very fast algorithm for generating such a matrix. Other generating
algorithms for these check matrices were discussed 
in \cite{Zha}\cite{GBT04}\cite{GBT05}. However, these algorithms are not in 
polynomial time.

It is obvious that if $m\cdot J$ is divisible by $R$,
the procedure of "placing a matrix upside down"
can be ignored. So,

{\bf Corollary 1 } Let $m\cdot J$ be divisible by $R$.
Suppose that $\Delta_{1}(R-1, J-1, m_{1})$ and
$\Delta_{2}(R-1, J, m-m_{1})$ are matrices with equal-weight-columns
and equal-weight-rows,
where $m_{1}=
[\frac{m\cdot J}{R} + \frac{R-1}{R}]$.  Then,

\begin{equation}
  \Delta'(R, J, m) = [ <1>(m_{1}) \oplus <0> (m-m_{1})]
                      \circ
          [\Delta_{1}(R-1, J-1, m_{1}) \oplus \Delta_{2}(R-1, J, m-m_{1})].  
\end{equation}

\noindent is a matrix with equal-weight-columns and equal-weight-rows.

{\bf Corollary 2 } There is $O(m\cdot R)$ algorithm for generating $\Delta(R, J, m=C_{R}^{J})$

Based on the above discussion, we can obtain two algorithms below:

{\bf Algorithm A: the none-recursive algorithm}

$\begin{array}{ll}
 
\mbox{\bf Step 1} & \mbox{Check if  $\Delta(R, J, m)$ satisfies the $L$-condition.} \\

\mbox {\bf Step 2} & \mbox{Decompose  $\Delta(R, J, m)$ into $\Delta(R_{1} , J_{1}, m_{1})$,}\\
                  & \mbox{$\Delta(R_{2} , J_{2}, m_{2})$, ..., $\Delta(R_{n} , J_{n}, m_{n})$,}\\
                  & \mbox{where each $\Delta(R_{i} , J_{i}, m_{i})$,(i=1,2,...,n) is a ending-state.}\\

\mbox {\bf Step 3} & \mbox {Merge $\Delta(R_{i} , J_{i}, m_{i})$,($i=1,2,...,n$)}\\ 
                   & \mbox {using $\overline{\oplus}$ operator.} \\
                   & \mbox {Note: "decomposition" and "merge" use the same objects.}
\end{array}$

{\bf Algorithm A': the recursive algorithm}

$\begin{array}{ll}
 
\mbox{\bf Step 1} & \mbox{Check if  $\Delta(R, J, m)$ satisfies $L$-condition.} \\

\mbox {\bf Step 2} & \mbox{Decompose  $\Delta(R, J, m)$ into $\Delta_{1}(R-1, J-1, m_{1})$}\\
                   & \mbox {and $\Delta(R-1 , J, m-m_{1})$.}\\

\mbox {\bf Step 3} & \mbox {Merge $\Delta_{1}(R-1, J-1, m_{1})$ and }
                    \mbox {$\Delta(R-1 , J, m-m_{1})$ based on (6).}
\end{array}$

\section {A Better Algorithm for Generating  $\Delta(R, J, m)$   }

Even though Theorem 2 did not give an $O(R\cdot m)$ algorithm,
it  provides a great hint to design a (virtually) optimum algorithm.
The key is to keep more 1's rows
at the top of the matrix and to avoid putting more as it needed.

The following simple calculations can be used to improve the process.
We would be able to know how many rows have the extra 1's in each
of $\Delta_{1}(R-1, J-1, m_{1})$ and
$\Delta_{2}(R-1, J, m-m_{1})$ before the merge. After the merge,
the array will not change.

Suppose that $\Delta_{1}(R-1, J-1, m_{1})$ and
$\Delta_{2}(R-1, J, m-m_{1})$ are Matrices with equal-weight-columns/rows, where $m_{1}=
[\frac{m\cdot J}{R} + \frac{R-1}{R}]$.

First, calculate
\begin{equation}
  r_{1} = (J-1)\cdot (m_{1}) \pmod{R-1}, 0 \le r_{1} <  (R-1), 
\end{equation}
\begin{equation}
  r_{2} = J\cdot (m-m_{1}) \pmod{R-1}, 0 \le r_{2} <  (R-1). 
\end{equation}

If $(r_{1} + r_{2} > (R-1) )$, let $r'= r_{1} + r_{2} - (R-1)$. Because
$r_{1} <  (R-1)$ and $r_{2} < (R-1)$,   $r_{1}>r'$  and $r_{2} >r'$.
Shift $r_{2}-r'$ rows in $\Delta_{2}$
to the bottom to obtain a $\Delta_{2}'$. If $(r_{1} + r_{2} \le (R-1) )$,
then move the first $r_{2}$ rows
to $r_{1}+1$ to $r_{1}+r_{2}$, so
\begin{equation}
  \Delta'(R, J, m) = [ <1>(m_{1}) \oplus <0> (m-m_{1})]
                      \circ
          [\Delta_{1}(R-1, J-1, m_{1}) \oplus \Delta_{2}'(R-1, J, m-m_{1})]  
\end{equation}
\noindent is a matrix with equal-weight-columns/rows.

The time complexity of the above process is
\begin{equation}
   T(m) = T(m_1)+T(m-m_{1})+ O(m\cdot R).
\end{equation}

Using the same technique to analyze the average case of Quick-Sort \cite{Cor},  we have
\begin{equation}
  T(m)= R\cdot m \log{(m\cdot R)} = T(m)= R\cdot m \log{(m)}.
\end{equation}

For the detail analysis, we could view
$T(m)$ as the average time required by the randomized $m_1$. That means
$m_1$ could be any number between $1$ to $m-1$. So
\begin{equation}
   T(m) = (T(1)+T(m-1) + T(2) + T(m-2)+...+T(m-1)+T(1) + O(R\cdot m\cdot m))/(m-1),
\end{equation}
\begin{equation}
   T(m) = \frac{2}{m-1} \Sigma_{k=1}^{m-1}T(k) + O(R\cdot m).
\end{equation}

\noindent Using the method of the mathematical induction, we can prove that 
$T(m)\le R\cdot m \cdot \log (R\cdot m)$ \cite{Cor}.

\noindent We know that $R\le m$ is held in most of cases.
Therefore, we have:

{\bf {Theorem 3}} The time complexity of generating  $\Delta(R, J, m)$ is
     $O( R\cdot m (\log R + \log m))$ for average cases, 
     or $O( R\cdot m \cdot \log m)$ for  most cases.

\section {More Efficient Algorithms for Generating  $\Delta(R, J, m)$?   }
                      
To avoid using extra time to build $\Delta(R, J, m)$,
we have to know exactly where to put the extra 1's if $r= J\cdot m \pmod{R}$
is not zero. Except in the case of $m=C^{J}_{R}$, even though
$r= J\cdot m \pmod{R}$, we may  have an
 unbalanced number of 1's at the next level.  For example,
 let $R=6$, $J= 3$, and $m =10$. $m_{1} = [(J\cdot m)/R + (R-1)/R] = 6$
and $(m-m_{1})=6$. $r = (J-1)\cdot m_{1})\pmod{(R-1)} = 12 \pmod {5} =2$.
Therefore, it seems impossible
to design an
$O(R\cdot m)$ algorithm for generating $\Delta(R, J, m)$. 
 In other words, it is reasonable to say that the best algorithm should
 use the divide-and-conquer to reach the
time complexity  $O(R\cdot m \log {m})$. Moreover,
to set up a dividing point in the middle of the length of the array might
break the $L$-condition, it seems hard to design an algorithm that makes
a matrix based on two equal sized arrays without the $L$-condition.
We have provided an analysis of average cases in Section 4.
We have the reason to believe that the algorithms discussed in this note
attain the optimal in theory. 
However, practically one can still develop more efficient algorithms for
generating such a matrix.

\section {Conclusion  }

This note translates the major steps of the author's earlier
paper written in Chinese for
optimal Hsiao codes. An algorithm analysis error was found and corrected.
The original algorithm has been modified and improved. A brief analysis
shows that the algorithm reached optimum in average cases if a
divide-and-conquer technique was involved in the algorithm.

{\it Acknowledgements} The author would like to present 
thanks to Dr. W. Stallings and Dr.
Nur Touba for their interests in his research that motivated 
the author to write this note.\\

\vskip 1in

\bf {\large Appendix }

\noindent ---------------------------------

\noindent Zbl 0629.65046

\noindent Chen, Li

\noindent An optimal generating algorithm for a matrix of equal-weight columns and quasi-equal-weight rows.
(Chinese. English summary)
[J] J. Nanjing Inst. Technol. 16, No.2, 33-39 (1986). [ISSN 0254-4180]

\noindent According to a recursive matrix which is introduced in this paper, an algorithm for a matrix of
equal-weight columns which are inequal each other and quasi-equal-weight rows is given. The algorithm
is optimal in time and space, and it can be applied to generate check matrices
 of optimal minimum
odd-weight-column SEC-DED codes. Finally, this algorithm is extended to a set of n elements.

\noindent MSC 2000:
*65F30 Other matrix algorithms

\noindent Keywords: recursive matrix; equal-weight columns; quasi-equal-weight rows; optimal minimum
odd-weight-column SEC-DED codes

\noindent -----------------------------------

\end{document}